# Realizing a Low-Power Head-Mounted Phase-Only Holographic Display by Light-Weight Compression

Burak Soner, Erdem Ulusoy, A. Murat Tekalp and Hakan Urey

***Abstract*— Head-mounted holographic displays (HMHD) are projected to be the first commercial realization of holographic video display systems. HMHDs use liquid crystal on silicon (LCoS) spatial light modulators (SLM), which are best suited to display phase-only holograms (POH). The performance/watt requirement of a monochrome, 60 fps Full HD, 2-eye, POH HMHD system is about 10 TFLOPS/W, which is orders of magnitude higher than that is achievable by commercially available mobile processors. To mitigate this compute power constraint, display-ready POHs shall be generated on a nearby server and sent to the HMHD in compressed form over a wireless link. This paper discusses design of a feasible HMHD-based augmented reality system, focusing on compression requirements and per-pixel rate-distortion trade-off for transmission of display-ready POH from the server to HMHD. Since the decoder in the HMHD needs to operate on low power, only coding methods that have low-power decoder implementation are considered. Effects of 2D phase unwrapping and flat quantization on compression performance are also reported. We next propose a versatile PCM-POH codec with progressive quantization that can adapt to SLM-dynamic-range and available bitrate, and features per-pixel rate-distortion control to achieve acceptable POH quality at target rates of 60-200 Mbit/s that can be reliably achieved by current wireless technologies. Our results demonstrate feasibility of realizing a low-power, quality-ensured, multi-user, interactive HMHD augmented reality system with commercially available components using the proposed adaptive compression of display-ready POH with light-weight decoding.**

***Index Terms*— holography, augmented reality, displays, wearable computers, data compression**

## I. Introduction

DIGITAL holography (DH) targets a wide range of application areas, including augmented reality (AR) and telepresence [1]. Computer-generated holography (CGH) enables synthesis of arbitrary wavefields of light with diffraction-limited resolutions, devoid of accommodation-vergence conflicts, accounting for all human visual cues [2]. Hence, multi-user and interactive AR systems based on high resolution and wide field-of-view CGH video displays promise the ultimate 3D visual experience.

TV-like holographic video display systems are not commercially available today since the display hardware for

Manuscript received 25 March 2019; revised 24 October 2019; revised 19 December 2019; accepted 29 January 2020.

The associate editor coordinating the review of this manuscript was Junsong Yuan.

The authors are with the Department of Electrical and Electronics Engineering, Koc University, 34450 Istanbul, Turkey. (e-mail: bsoner16@ku.edu.tr)

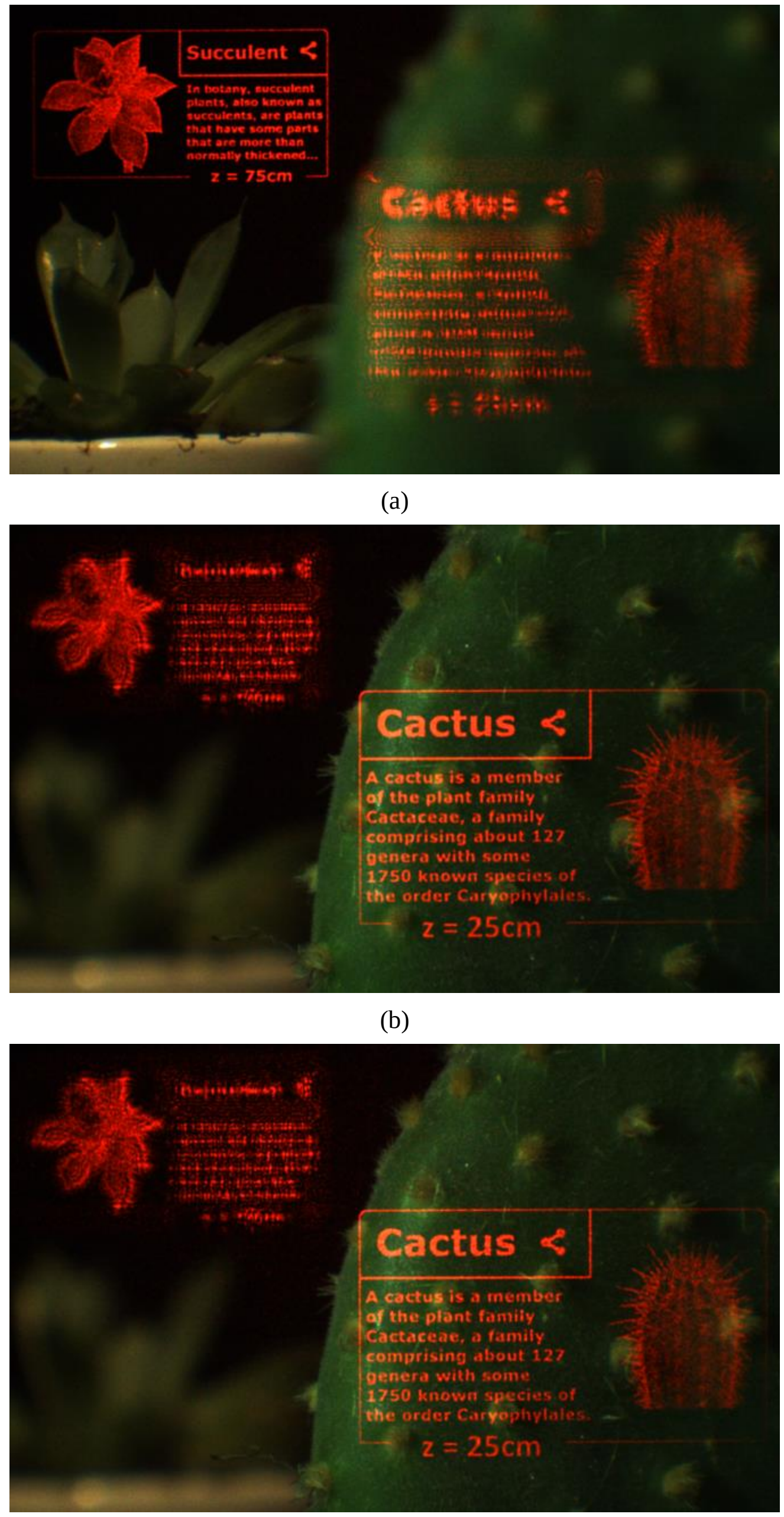

(a) (b) (c)

Fig. 1. Pictures of actual phase-only holograms captured by a camera in our lab showing two "info card" objects at different depths overlaid on real objects to demonstrate an AR use case. Uncompressed CGH with the camera focused on "Succulent" at 75cm is shown in (a) and focused on "Cactus" at 25cm is shown in (b). The latter CGH compressed at 3 bpp with uniform PCM is shown in (c). This result demonstrates acceptable CGH quality can be obtained at 3 bpp; however, the resulting bitrate per second at 60 fps exceeds that needed for reliable communication with current wireless technologies.

such a system would require on the order of Tera sub-micron sized pixels addressable at video rates for satisfactory field-of-view (FoV) and display quality. Challenges associated with building such high-space-bandwidth-product display hardware and with the efficient generation and transmission of CGH for such displays are investigated in detail in [3]. On the other hand, head-mounted holographic displays (HMHD) only provide content to the users' eye-box rather than the whole FoV, significantly reducing the bandwidth requirement. HMHDs are thus projected to be the first commercial realization of holographic video display systems [3].

HMHDs utilize spatial light modulators (SLM) for displaying holograms, generally one SLM per eye [4]. Since holograms are complex-valued wavefields, an ideal SLM should modulate both the amplitude and phase of the light. Although research for building "complex-mode" SLMs is ongoing [5], such devices are not commercially available today. Phase-mode SLMs, which can display phase-only holograms (POH), have higher diffraction efficiency compared to amplitude-mode and complex-mode SLMs. Full HD (1920 x 1080 pixels), 60 fps, phase-mode SLMs with 8-bit/2π dynamic modulation range have been demonstrated to provide acceptable visual quality as near-eye holographic displays [4]. Due to their proven performance, we focus on HMHDs using Full HD, 60 fps phase-mode SLMs with 8-bit/2π modulation range in order to display monochrome POHs.

Generating acceptable quality 60 fps Full HD CGHs typically require powerful GPU-based workstations capable of ~20TFLOPS in >200W power [6]. A standalone CGH-generating HMHD unit would need to achieve this at a power budget of <5W [7]. Since FLOPS/W figures of current mobile processors are orders of magnitude away from this requirement [8], CGHs need to be generated on a server workstation and transmitted to the HMHD, leaving only the pose/gaze estimation, CGH decoding and display tasks of the hologram pipeline to the HMHD. Such a system is depicted in Fig. 2. Furthermore, current wireless communication technologies cannot reliably achieve the transmission rate required for a monochrome, 60 fps Full HD, 2-eye CGH, which is on the order of a few Gbit/s. Therefore, the CGH needs to be compressed before transmission to the HMHD. Section II discusses these requirements in detail.

There have been two main approaches for compression of static holograms in the literature. They are coding data in: (1) the hologram plane and, (2) an intermediate plane which can be transformed/propagated to the hologram plane after decoding. The authors in [3] further divided the second category into coding of the input content, of intermediate time-frequency representations of holograms via nonlinear canonical transforms [9], and coding in the backpropagated object plane, with detailed explanations. Category (2) of abovementioned compression methods will not be analyzed in this paper since, similar to the standalone CGH generation task, they would raise the FLOPS/W requirement to currently unrealizable levels.

Prior work for category (1) builds on top of existing 2D image compression methods recognizing the fact that statistics of holograms are different from natural images. A thorough review and classification of prior art and the current state-of-the-art, ranging from modifications on existing codecs [11-15] to wavelet-based methods specifically designed for holograms [16-20] and to view-dependent, content-aware methods [21-23], is presented in [3] and [10]. We further characterize relevant codecs from these works in Section III. Majority of prior work on category (1) considered hologram data as complex wavefields composed of either real-imaginary or amplitude-phase components, or as an intensity-based representation (real-valued representations like phase-shifted distances or holographic recordings with imaging sensors [10]). However, if complex hologram coding was employed for HMHDs that use phase-mode SLMs, the decoded complex hologram would need to be transformed into a POH in the HMHD mobile computer prior to display. This task, like the standalone CGH generation task, would raise the FLOPS/W requirement of the HMHD computer to currently unrealizable levels. Hence, "display-ready" POHs must be compressed and transmitted to HMHDs, where they will merely be decoded and displayed as further discussed in Section II.

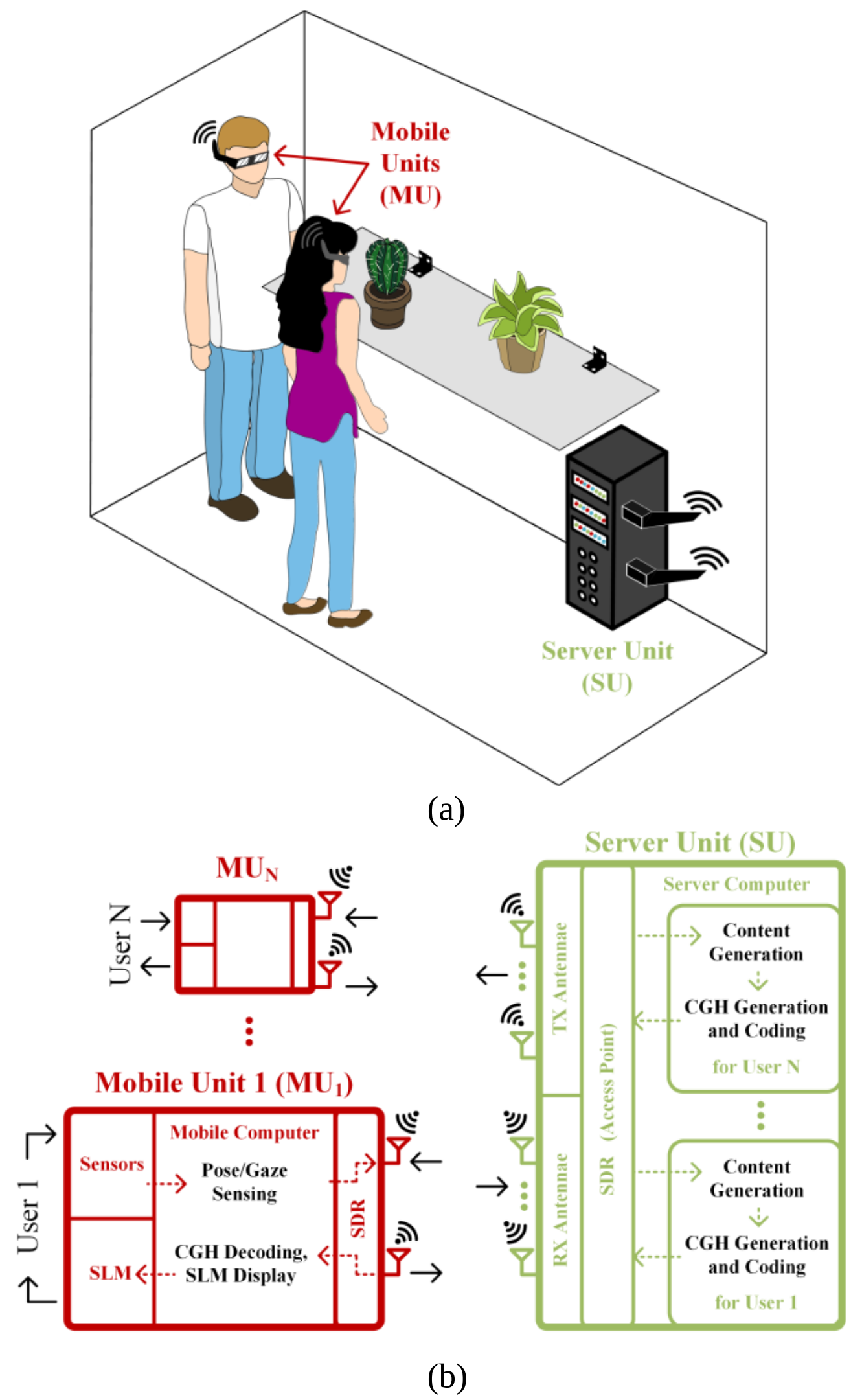

Fig. 2. (a) Depiction of the example AR application scenario demonstrated in Fig. 1. (b) The system architecture for a realizable HMHD-based AR system, which serves multiple users from a single server unit. All MUs are identical.

High quality POHs can be generated by applying Gerchberg-Saxton (GS) type iterative algorithms to the complex hologram [24]. Other non-iterative methods, such as error diffusion or simply discarding the amplitude, are not suitable for display applications since the result suffers heavily in either quality or resolution [6]. Note that POH coding is not equivalent to coding the phase of the complex hologram. Authors in [25] presented an extensive benchmark for coding of complex holograms and their associated phase values. POHs have additional characteristics inherited from the transformation method (e.g., a uniform distribution of phase samples for GS-type iterative methods, as shown in Fig. 3). Hence, re-evaluating the coding methods previously applied to complex holograms for "display-ready" POH coding, is also necessary. Section III analyzes available compression methods that are applicable to POHs obtained via the GS-type iterative methods since they provide the best uncompressed hologram quality.

The main contributions of this paper are:

- We propose a realizable system for multi-user and interactive HMHD-based AR applications. Identifying the main design requirements, compression and wireless transmission of "display-ready" POHs emerge as a key requirement.
- We analyze the performance of existing light-weight compression methods for compression of POHs generated by GS-type iterative methods.
- We propose a new low-power versatile POH codec that performs progressive quantization and pixel-wise rate-distortion control, enabling feasibility of transmission of display-ready POH over current wireless LAN channels in multi-user scenarios.

An application for this system is demonstrated in Fig. 1, where two "info card" objects at different depths are displayed via a monochrome CGH (generated with the RGB+Depth method [26]) on a prototype HMHD. While Fig. 1(a) and (b) demonstrate the multi-depth capability, Fig. 1(c) demonstrates acceptable quality at 3 bpp compression via uniform pulse code modulation (PCM). Our proposed codec further lowers the rate to enable utilization of current wireless LAN channels.

The rest of the paper is organized as follows: the proposed HMHD-based AR system design requirements are presented in Section II. Performance of existing POH compression methods are analyzed in Section III. A new versatile PCM-POH codec for HMHDs with progressive quantization and pixel-wise rate-distortion characteristics is proposed in Section IV. Section V concludes the paper by emphasizing the feasibility of our design methodology for HMHD-based AR systems using off-the-shelf components and the proposed low-power PCM-POH codec.

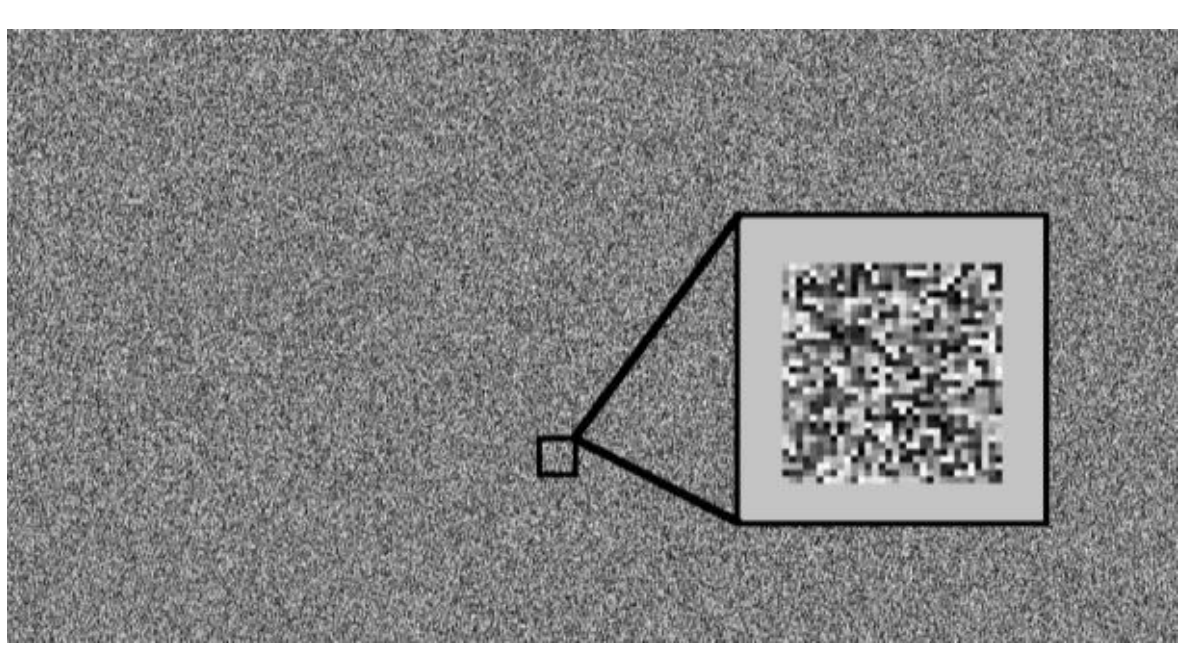

Fig. 3. Example POH computed with a Gerchberg-Saxton-type iterative method. In addition to high spatial frequencies, phase samples are uniformly distributed. The inset depicts a 32x32 patch to illustrate signal characteristics.

## II. Design Requirements for HMHD-based AR System

This section identifies the key design constraints and requirements for an HMHD-based AR system architecture that is feasible utilizing commercially available mobile processors, Full HD, 60 fps phase-mode SLMs to display POH, and current wireless communication technologies.

With an operational requirement of ~3 hours on batteries, the HMHD has a power budget of 5W at best [7]. Considering that a fraction of this budget is available for computation and that commercially available mobile processors can achieve maximum ~200 GFLOPS/W [8], standalone hologram generation (~10 TFLOPS/W) or even transformation of complex holograms to POH cannot be realized on the HMHD. This stringent power constraint (PC) necessitates that these computations be offloaded to a server unit.

Considering wireless transmission of the results of offloaded computations back to the HMHD, the system runs into a rate constraint (RC) since current wireless communication technologies cannot support the required multi Gbit/s rates. IEEE 802.11ac [27], currently the most reliable high performance wireless local area network (WLAN) standard, can support 80MHz of bandwidth and a 16-QAM modulation order, translating to 60-200 Mbit/s rates per channel, under realistic channel conditions [28]. Therefore, RC necessitates that the results of offloaded computations are compressed to at least this bitrate range prior to wireless transmission.

Under influence of these dominant constraints, the following design requirements (*R#*) emerge for a feasible architecture:

- *R1:* Hologram generation needs to be offloaded to a server which wirelessly transmits the result back to HMHD since HMHD cannot achieve this task due to PC.
- *R2:* The generated hologram needs to be compressed on the server prior to transmission to HMHD due to RC.
- *R3:* The HMHD SLMs can only display POHs. Although complex holograms are more compressible than POHs, transformation of the complex hologram to its associated POH cannot be realized on the HMHD due to PC. Therefore, compression should be applied on the "display-ready" POH rather than the complex hologram.
- *R4:* While current WLAN channels can reliably sustain about 60-200 Mbit/s, on average this is nearly 1/20$^{th}$ of the uncompressed POH rate, 1.85 Gbit/s. Therefore, the POH codec should provide a highly versatile rate-distortion control mechanism for flexibility in congested multi-user scenarios (RC).
- *R5:* The POH decoder on the HMHD needs to be low-power due to PC. Therefore, only low-complexity decoders or proven codecs with low-power ASICs that are commercially available, can be used.

While *R1* is self-explanatory, *R2* and *R3* signify the need for a display-ready-POH codec and *R4* and *R5* express certain requirements for the realization of this codec. The proposed feasible system architecture, which satisfies all five requirements, is shown in Fig. 2(b).

To comply with *R4*, the codec needs to be aware of current wireless channel conditions, such as congestion, signal-to-noise-ratio (SNR) etc. and thus determine the maximum rate per channel at which reliable communication can be sustained. To compress the hologram at this rate, it should then exploit the redundancy due to SLM hardware limits and content sparsity, and employ region-of-interest coding. In order to ensure it has low latency access to such information, the POH codec should be realized on the co-processor of a low size weight and power (SWaP) software defined radio (SDR), which handles the wireless transmission. Such low SWaP (credit-card sized, mini-PCIe interface, ~1W [29]) SDRs are commercially available, with usually FPGAs as co-processors. Realizing the codec on an SDR rather than on a dedicated hardware radio enables complete control over compression and transmission with low-latency connection between them.

In order to resolve *R5* with a codec that can comply with *R4* subject to the power constraint (PC), Section III analyses POH compression using existing codecs for which there are low-power codec ASICs available, such as BPG (HEVC-Intra), JPEG 2000 Part I (JP2K-I) and JPEG, and simple direct pixel quantization (PCM).

## III. Analysis of Available POH Compression Methods

Methods that can be considered for compression of POH data can be grouped as standard image codecs, complex wavefield codecs, and special methods as shown in Table I. There exist low-power image compression ASICs that are rated at less than 0.2nJ/px [33] for standard codecs such as JPEG, JP2K-I and HEVC-Intra, which are feasible to use in a low-power HMHD. On the other hand, only software implementations exist for complex wavefield coding methods that were identified in [3] as having high rate-distortion efficiency. These methods are: modified JPEG 2000 Part II (JP2K-II) implementations that either utilize directional-adaptive wavelets and full-packet decompositions [12] or wave atom transforms [34], and a modified HEVC-Intra codec using adapted transforms [24]. It is not feasible to implement the decoders of these methods in software in a low-power HMHD due to their high computational complexity. There also exist special methods for POH coding such as phase-difference-based-compression (PDBC) [15]. The PDBC decoder requires both HEVC-Intra and JBIG decoders and additional soft components to merge their outputs and therefore it also is not feasible for implementation in a low-power HMHD. Alternatively, we can use PCM for POH coding. PCM has multiple orders of magnitude less computational complexity compared to these other decoders, which makes its low-power soft implementation feasible in our proposed system.

Hence, this section analyzes the performance of codecs that meet the requirement *R5* stated in Section II. They include HEVC-Intra (BPG) [30], JPEG2000 [31], JPEG [32], which have low-power ASICs and the PCM method, which has a very simple software decoder that is feasible for use in a low-power HMHD. Performance of the special POH coding method, PDBC [15], is also analyzed. Effects of flat quantization of transform coefficients and phase unwrapping on compression performance are investigated. We evaluate three main approaches in the following subsections:

1) Apply standard codecs and flat quantization (flat-q)
2) Apply standard codecs + flat-q + phase unwrapping
3) Apply special methods for POH coding (PCM and PDBC)

### A. Standard Codecs and Flat Quantization

HEVC-Intra is currently the state-of-the-art lossy image coder, derived from the intra-frame coder of the high-efficiency video coding (HEVC) standard. BPG is an open source implementation of this codec. JP2K-I uses wavelet transforms while JPEG and BPG uses cosine transforms on square blocks from the image (8x8 pixels for JPEG, variable block size 4x4 – 64x64 for BPG) specifying the granularity of quantization steps applied to each of the transform coefficients within that block with quantization matrices. Since JPEG and BPG were designed for natural images/photographs, which dominantly have lower-frequency content, their default quantization schemes favor lower frequencies (transform coefficients). As can be seen from the inset in Fig. 3, sub-blocks in a POH are nearly full-band signals; thus, it is evident that JPEG and BPG with flat quantization would perform better on POH. To demonstrate this, both schemes were evaluated. The 8-bit versions of the codecs were used in this class of compression.

TABLE I
METHODS CONSIDERED FOR COMPRESSION OF POH DATA FOR AR APPLICATIONS WITH HMHDS

| Category | Compression Method | Low-power decoder available in the form of | |
|---|---|---|---|
| | | commercial ASIC | software realization [a] |
| Standard image codecs | JPEG [30] | ✓ | |
| | JP2K-I [31] | ✓ | |
| | HEVC-Intra (BPG) [32] | ✓ | |
| Complex wavefield codecs | JP2K-II DA-DWT [12] | | |
| | JP2K-II WA [34] | | |
| | HEVC-Intra AT [25] | | |
| Special methods for POH coding | PDBC-HEVC-Intra [15] | | |
| | PCM | | ✓ |

[a] Computational complexity is very low; therefore, software realization on a low-power processor is feasible.

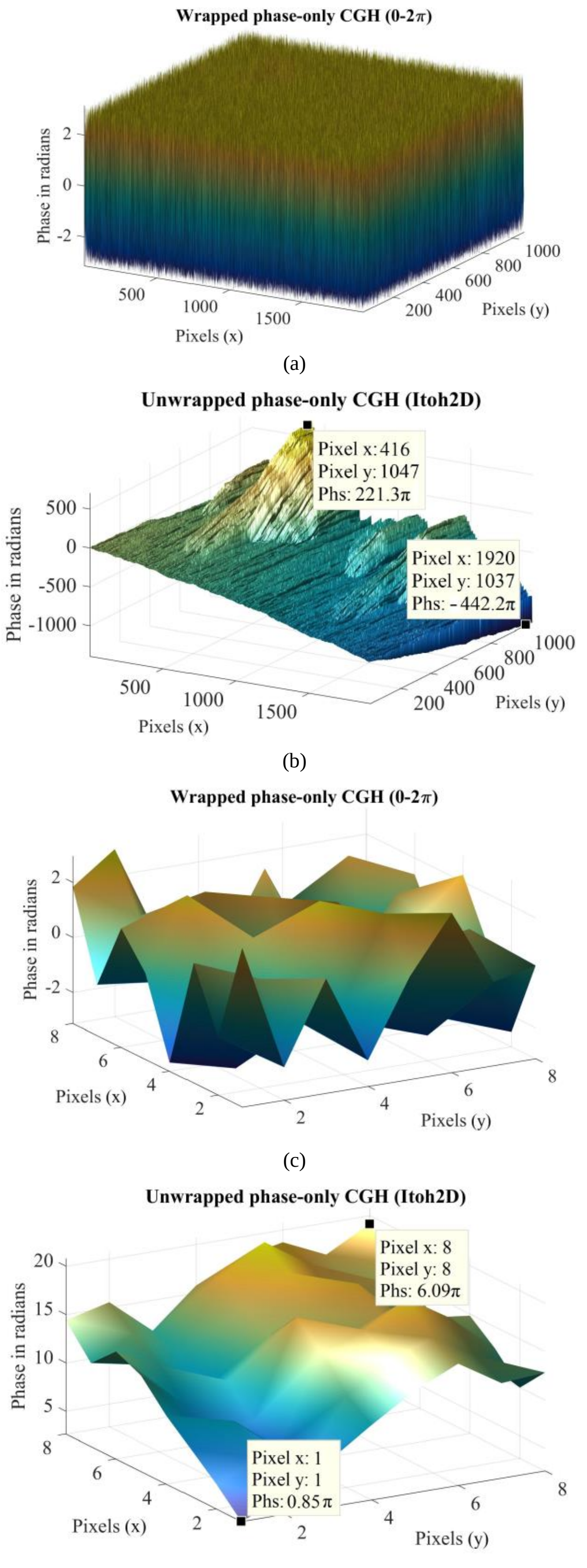

(a) (b) (c) (d)

Fig. 4. A 1920x1080 POH wrapped to ±π in (a) and unwrapped in (b). Phase range required in (b) is ~2084 radians, ~663.5π. To completely represent the unwrapped version while avoiding phase resolution loss with respect to the 8-bit, 2π representation, a 17-bit (1024π) codec would be necessary. An 8x8 block from (a) is shown in (c). Applying the same operation produces the much smoother signal in (d) and a phase range of ~16.4 radians, 5.2π which allows a 10-bit (8π) codec without loss.

### B. *Does Phase-unwrapping Help Compression?*

Phase unwrapping removes the unwanted jumps from 2π-wrapped 2D phase data to reconstruct the desired smooth phase information. Since each 8-bit pixel of the display-ready POH represents a phase shift between 0-2π (by 0-255) applied by each SLM pixel, the POH can be preprocessed for phase unwrapping to achieve smoother, possibly more compressible data. Naturally though, the resulting unwrapped data is not bound to the 0-2π dynamic range anymore so there is a tradeoff between an increase in resulting raw data size and data compressibility. A detailed description of the Itoh 2D phase unwrapping algorithm used in this study can be found in [35].

Phase unwrapping has more potential for the hologram of a true phase object rather than a POH, which is merely the product of an iterative optimization scheme trying to best represent a complex wavefield with only phase information. Nevertheless, phase unwrapping is a lossless, smoothing pre-processing step which POH compression can benefit from.

2D phase unwrapping for compression can be implemented in two ways: 1) unwrap the whole POH and compress, or 2) unwrap blocks within the POH independently and compress. Either way, the number of bits representing the range of unwrapped phase needs to be increased to avoid phase resolution loss. Suppose that the wrapped phase samples in 0-2π are represented with 8-bits/sample. If the unwrapping results in phase samples between 0-8π, the resulting representation should use 10-bits/sample since there is a factor of 4, corresponding to two additional bits. This issue has been addressed by using higher dynamic range options of BPG (10-, 12-, 14-bit), JPEG (12-bit) and JP2K-I (16-bit).

The phase range increase due to unwrapping the whole POH is very large and it scales up with resolution. An example wrapped 1920x1080 POH and its unwrapped version are shown in Fig. 4. The phase range required is ~663.5π. The only codec option that could meaningfully represent such a large range was the 16-bit JP2K-I, which allows a 0-512π range without phase resolution loss. Unwrapped POHs which exceeded this range were wrapped in modulo 512π to avoid resolution loss.

Block sizes for block-unwrapping were chosen in conjunction with BPG and JPEG transform block sizes. The tiling option in JP2K-I was not exploited in this study since it did not provide small enough block sizes, causing the phase range increase (JP2K-I only has 8- and 16-bit, so it is a x2 increase) to null out the compressibility gain. A block size of 8x8 was chosen both since it matches the JPEG and BPG default block size options and since the range increase with 8x8 was mostly confined to the 0-8π range, enabling the BPG 10-bit and JPEG 12-bit without phase resolution loss. To see the effect of phase resolution loss within display quality limits, BPG 8-bit for the 0-4π range was also evaluated.

### C. *Special Methods for POH Coding*

The phase samples of POHs obtained via GS-type iterative methods investigated in this paper are uniformly distributed. Since simple scalar quantization works best for compressing uniformly distributed data [36], PCM was applied to POHs.

In the case of no compression, each POH pixel carries 8-bit phase samples (256 quantization steps) for the 8-bit SLM. In the case of PCM compression, the number of represented quantization steps are decreased but the values still address the 0-2π range with an 8-bit container since that is what the SLM expects (e.g. a 2.322 bpp output from PCM uses $\log_2(2.322)$ = 5 steps, which are 0, 64, 128, 191 and 255). The number of symbols getting smaller constitutes the compression.

Note that PCM dictates a predetermined number of quantization step variations (i.e., 256) and thus, predetermined corresponding bitrate values, increasing in logarithmic fashion. The smallest bitrate possible is 1 bpp, which corresponds to a binary hologram. Although in common image compression schemes, bitrates much lower than 1 bpp are possible with acceptable quality, the methods mentioned above for POH compression have not been observed to produce acceptable quality reconstructions for such bitrates. Therefore, PCM is also a viable option for POH compression.

Since PCM is basically scalar quantization, a rebuild of the BPG codec which skips the inherent cosine transform, called "BPG transform-skip", was also evaluated. Since this rebuild is basically PCM + BPG lossless coding tools which normally code transform coefficients, this method was evaluated for its potential of enhancing BPG and surpassing PCM performance.

PDBC [15] codes the phase distance image (i.e., absolute value of differences from a reference pixel) which is more compressible than the original but requires the transmission of the associated JBIG-coded binary sign image. The HEVC-Intra based implementation of this strategy from [15] was chosen for the analyses in this section since it gives the best results.

### *D. Evaluation*

The subject of evaluation in this study is POHs of content that is suitable for AR. Since holographic AR content is intended to be overlaid on real objects which are viewed through a beam splitter, text, symbology, computer-generated characters and images on a black background are of interest. The physical spaces corresponding to the black parts in holographic content become see-through on the physical display. For this reason, evaluation scenarios in this study were confined to POHs of such grayscale content. 15 of the black-background portraits from photographer Nelli Palomäki [37] and 1 black-background photograph courtesy of Stock Footage, Inc. [38] were used alongside numerous appropriate mixed symbology/grayscale content generated by the authors.

#### *1) Quantitative evaluation methodology*

Quantitative evaluation of compression performance on "display-ready" POHs in this study is based on numerical optical reconstructions which mimic the physical reconstruction procedure. The effects of optical components and the SLM on the coherent, unmodulated illumination are simulated numerically and the intensity of the resulting complex wavefield, which corresponds to what the viewer sees on the actual setup, is produced. Often, a hologram contains information at multiple depth planes, but the user focuses his/her eye at one of them at a given time, blurring out the peripheral from his/her point of view (e.g. succulent and cactus in Fig. 1(a)). Previous research has shown peripheral quality in in a near-eye setup has a much smaller effect on the overall quality perception compared to that of the foveal region (where the user focuses on) [39]. Therefore, this study has concentrated on content where the whole frame is kept at a single depth, the whole frame corresponding to the foveal region.

The evaluation pipeline works as follows. The complex Fresnel CGH [6] is generated for a certain focus depth using the RGB+D method [26]. Afterwards, the associated POH is obtained using a GS-type iterative algorithm, namely Fienup with Don't Care Regions (FIDOC) [24]. This POH, with 8-bit phase samples in the 0-2π range in each pixel, is then numerically reconstructed at the viewing plane for that focus depth, to obtain the "best", i.e., uncompressed, reconstruction.

In parallel, this POH is compressed via the methods (*M#*) listed in Table II. The de-compressed POH is then reconstructed and compared to the best reconstruction with the peak-SNR (PSNR) quality metric. The result is a PSNR vs. bpp array for the CGH of the given content at a certain focus depth, compressed with one of the methods specified in Table II. Since different focus depths produce different POHs, for each content, this procedure is repeated for depths 25-500cm, which constitutes the meaningful human visual range. Mean and standard deviation with respect to depth are recorded for each data point, constituting the overall compression performance result for that given content, with the specified compression method. This evaluation is then repeated for different grayscale AR content samples to further ensure statistical significance.

#### *2) Experimental setup*

For experiments, a Holoeye-Pluto SLM (8 μm pixel pitch, 8-bit pixel phase modulation depth) was illuminated by a collimated 638nm HeNe laser and the modulated beam was directed with a pellicle beam splitter towards an aperture and imaging lens. A FLIR Flea3-USB camera was used for capture.

The numerical reconstructions basically assume an ideal SLM which can do 8-bit, 2π modulation, a perfectly uniform and coherent wavefront illuminating the SLM and perfectly aligned optics. Due to setup defects like dirty components, minor misalignments in optics, fiber-coupled laser beam imperfections and a non-ideal SLM, the "best numerical reconstruction" is nearly impossible to obtain experimentally. Without a comparable best reconstruction on the experimental side, a PSNR vs. bpp comparison of numerical vs. experimental results would not give correct results. For this reason, the experimental validation of numerical reconstructions is done qualitatively by viewing the artifacts and visible quality degradations. An example result is shown in Fig. 7.

### *E. Results and Discussion*

For easier interpretation of the results from the 10 methods mentioned in Table II, 3 evaluation groups were created. These groups are described in Table II. The results for evaluation groups 1, 2 and 3 can be seen in Fig. 5(a), (b) and (c) respectively. The content used for these results was the one in Fig. 1 with the cactus and succulent info card objects.

TABLE II
QUANTITATIVE EVALUATION METHODS

| Evaluation Group | Relates to | *M#* | Codec Info | Quantization Scheme | Phase Unwrapping? | Codec Dynamic Range |
|---|---|---|---|---|---|---|
| Group 1 | Section III-A | *M1* | JPEG | Default | No | 8-bit/2π |
| | | *M2* | JPEG | Flat | No | 8-bit/2π |
| | | *M3* | JP2K-I | - | No | 8-bit/2π |
| | | *M4* | BPG | Flat | No | 8-bit/2π |
| Group 2 | Section III-B | *M5* | JP2K-I | - | Total | 16-bit/512π |
| | | *M6* | JPEG | Default | 8x8 Block | 12-bit/32π |
| | | *M7* | BPG | Flat | 8x8 Block | 10-bit/8π |
| | | *M8* | BPG [a] | Flat | 8x8 Block | 8-bit/4π |
| Group 3 | Section III-C | *M9* | PCM | - | No | 8-bit/2π |
| | | *M10* | BPG [b] | - | No | 8-bit/2π |
| | | *M11* | PDBC-HEVC-Intra | Default | No | 8-bit/2π |

[a] *M4* represents a 2π phase range with 8-bits, *M8* incurs phase resolution loss since it tries to represent a larger range with same number of bits.
[b] *M10* is the BPG transform-skip method mentioned in Section III-C.

As mentioned in Section III-D-1, in order to evaluate foveal hologram quality, both objects were kept at the same focus depth, rather than one far and one close like in the Fig. 1. All curves show the mean result where a standard deviation of ~2-3 dB with respect to foveal focus depth (samples from iteration over the depth range 25-500cm, as mentioned in Section III-D-1) was observed for each curve. Trials with the other content mentioned at Section III-D resulted in similar relative performance between different methods. While sparser content gave better compression performance (up to 8 dB improvement at same bpp for content with meaningful sparsity), resolution was not observed to be a major factor.

In group 1, BPG with flat quantization (*M4*) was the best performer. Since BPG with flat quantization outperforms default BPG, only results for the flat version are included. As seen in group 2, phase unwrapping does not boost compression performance and *M4* is better than the best contender, the 10-bit, 0-8π BPG on 8x8 block-unwrapped POH (*M8*). The increase in compressibility due to phase unwrapping cannot overcome the data size increase, resulting in a decrease in compression performance due to its use. In group 3, PCM (*M9*) performs slightly better than *M4*, BPG transform-skip (*M10*) and PDBC-HEVC-Intra (*M11*) especially for >25dB PSNR, which is observed as the acceptable quality range in experiments (Fig. 7). We next propose a PCM-based POH codec to achieve realizable rates since PCM is the best performer among codecs that satisfy the power constraint *R5*.

## IV. A LOW-POWER VERSATILE PCM-POH CODEC

Based on results from Section III, this section proposes a new PCM-based POH codec that provides a flexible rate-distortion control mechanism for the HMHD-based AR system, exploiting the sparsity of the content, to make it compliant with the feasibility requirements *R4* and *R5* mentioned in Section II. Utilization of PCM enables 2 main features:

- Progressive quantization, which supports SLM-dynamic-range dependent quantization
- Per-pixel rate-distortion control

### A. Progressive Quantization

Since PCM simply applies scalar quantization, a progressive quantization scheme can be employed for a POH codec employing PCM. The more significant bits (number determined by the maximum reliable communication rate available on the channel at the time) can be sent first, and less significant bits can follow in subsequent packages, ensuring a certain quality of service (QoS) which increases in a progressive manner with available bandwidth. An example with 2, 3, 8 bpp progressive quantization is shown in Fig. 7.

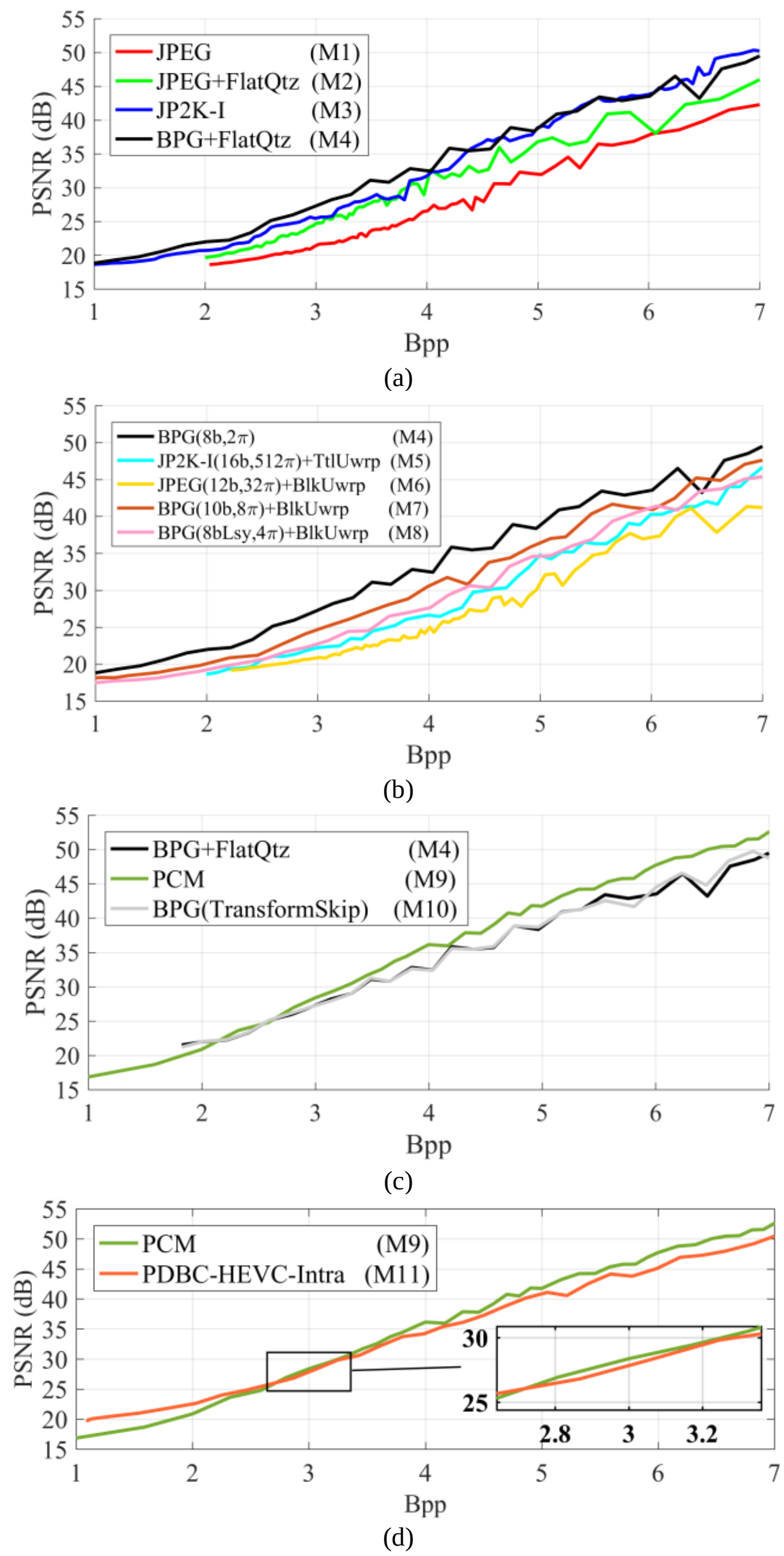

Fig. 5. Results for evaluation groups 1, 2 and 3 defined in Section III-E are shown in (a), (b) and (c)-(d) respectively. BPG+FlatQtz is the best in the first 2 groups. For the acceptable quality range of about >25dB PSNR, PCM performs better than other special methods for POH coding in group 3 by a small margin. PCM is the best method that satisfies constraints *R4* and *R5* from Section II.

SLMs cannot fully realize their advertised 8-bit/2π dynamic modulation range and lose the less significant bits of the hologram pixels, causing a practical upper bitrate limit for the progressive quantization scheme over which the display quality does not increase any further. For the SLM used in this study, an effective dynamic modulation range of ~4-bit/2π was observed during experimental evaluations. This reduction in the dynamic range of the SLM is due to the fluctuations (mainly due to temperature, the target “gray” level and drive waveform shapes) in the liquid crystal cell voltages [40]. Since PCM effectively reduces the resolution in modulation depth to compress data, the codec directly exploits this redundancy.

### B. *Per-pixel Rate-Distortion Control*

Most state-of-the-art 2D image/video codecs employ rate-distortion optimization and control over regions of interest (RoI) inside the data. These codecs use regions (a block of pixels) as their minimum addressable unit though and cannot offer truly arbitrary shaped RoI coding. For example, JPEG can alter the quality factor for 8x8 transform blocks and BPG has similar, more sophisticated features, but BPG also works on block-based RoI. A codec utilizing PCM has the advantage of addressing individual pixels on the subject data for rate-distortion control, providing the ultimate flexibility.

A per-pixel rate-distortion control algorithm for holograms needs to address a fundamental distinction: unlike pixels on a conventional 2D display, a POH pixel on the SLM carries information from multiple object points. Since the Fresnel CGHs investigated in this study are computed via virtual light propagation from each object point with respect to a finite aperture (i.e., eye-box), each point contributes to regions on the POH rather than individual pixels, where the region size changes with object point depth. However, since HMHDs utilize small apertures compared to SLMs and object point depths are within the human visual range (25-500cm), these regions are small; typically, less than 1% of the SLM area [6]. A detailed discussion on this well-known phenomenon is provided in [6] where these regions are called “sub-holograms”.

For HMHDs, this phenomenon leads to the following: POH pixels that are far away from their associated object points/pixels in the original scene (i.e., not within the region to which that point contributes) do not contribute significantly to the appearance of that object on the final displayed hologram. Since AR application scenarios for the proposed HMHD system do not consider objects covering the total FoV or full scenes with backgrounds, and since the POH regions relevant to these objects are only about 1% larger than the objects themselves for the proposed system as discussed above, a significant portion of the POH pixels fall into this category. Accordingly, we have observed that up to ~70% of POH pixels can simply be not coded in commercially relevant application scenarios for the proposed system such as in Fig. 1 (where 65% of pixels can be not coded), without significant loss in quality. The remaining pixels constitute the RoI. The per-pixel rate-distortion control feature of the proposed PCM-POH codec directly exploits this redundancy since it accepts such arbitrary sized RoI and therefore allows for up to 3.33× near-lossless reduction in rate.

### C. *Achieving Realizable Rates with the PCM-POH Codec*

Each HMHD unit in the proposed AR system implements a rate control algorithm which decides on a desired rate-distortion performance utilizing the features of the PCM-POH codec given the maximum available reliable communication rate.

With the SLM-dynamic-range-aware feature, the rate distortion algorithm enables the codec to determine an upper limit on bit rate over which the CGH does not look any better on the HMHD than the uncompressed one. Next, depending on the object points in the scene, the rate distortion algorithm provides a RoI to the codec. Coding only the RoI efficiently utilizing the per-pixel rate-distortion control feature, our codec can achieve up to ~9× compression (obtaining 3.33× from RoI and 2.67× from acceptable quality 3 bpp RoI coding, as shown in Fig. 1 and Fig. 7), which brings us from the uncompressed POH rate of 1.85 Gbit/s down to the realizable 200 Mbit/s rate.

Furthermore, the available rate varies between 60-200 Mbit/s with respect to channel conditions as mentioned in Section II.

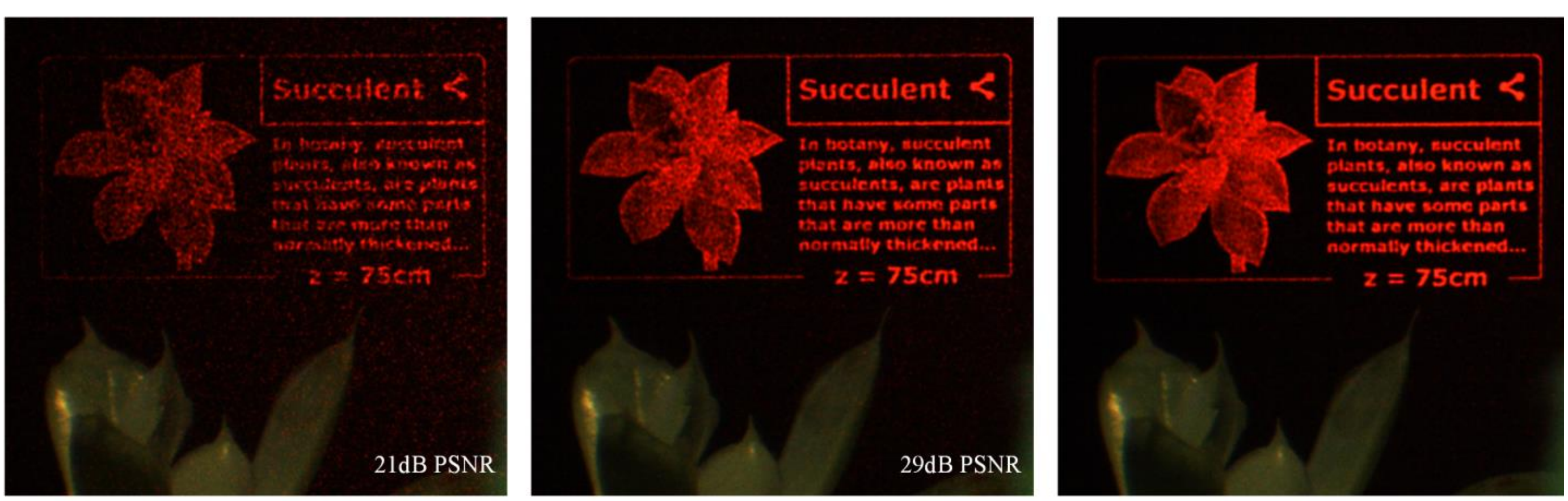

2bpp PCM-Compressed Reconstruction | 3bpp PCM-Compressed Reconstruction | Uncompressed Reconstruction

Fig. 7. POH reconstructions for the AR use case in Fig. 1 showing the visual effects of progressive quantization for 2 bpp, 3 bpp and 8 bpp (uncompressed) bitrate.

The proposed codec adapts to the channel conditions by employing progressive quantization up to the SLM-limit, providing a progressive increase in quality with respect to the available rate.

### *D. Serving Multiple Users*

The proposed HMHD system uses IEEE 802.11ac for wireless communication. 802.11ac has two main features that enables serving multiple users with high rates: simultaneous and non-interfering spatial streams (SS) and high-bandwidth channels. It supports 5 non-interfering 80MHz-wide channels around the 5GHz band in most countries [27] that can reliably attain ~60-200 Mbit/s rates each under realistic channel conditions [28]. While each user in the proposed system uses one channel, access points (AP) use SS on directional beams to reuse the channels for spatially separated users. Since 802.11ac supports 8 simultaneous SS, a capable 802.11ac AP can support reliable channels for $5 \times 8 = 40$ spatially distributed users. As depicted in Fig. 2(b), our system implements this "capable AP" via a high-end SDR on the server side and user HMHDs utilize low-SWaP SDRs capable of one SS and one 80MHz channel. Therefore, our system can support e.g., a classroom application scenario (Fig. 1) for up to 40 simultaneous users.

## V. Conclusion

Head-mounted holographic displays (HMHD) have a much lower space-bandwidth product requirement compared to TV-like holographic video display systems since they only need to provide data to the user's eye-box, and thus are projected to be the first commercial realization of holographic video display systems. HMHDs use phase-only LCoS SLMs which can only display phase-only holograms (POH) and are currently available for 60 fps Full HD. Performance/watt requirements for standalone hologram generation, transformation of complex holograms to POH, or even complex decoding on the HMHD are too high for commercially available mobile processors (which can achieve max. <200 GFLOPS/W), necessitating generation of "display-ready" POH on a nearby server and transmission to the HMHD after simple compression that would allow light-weight decompression. Two main design constraints for a realizable HMHD, namely power and transmission rate, and associated system design requirements were identified. We present a feasible architecture for a multi-user, interactive, 60 fps Full HD, monochrome HMHD-based augmented reality system under the design constraints and requirements, focusing our discussion on compression and the effective utilization of associated rate-distortion trade-offs for wireless transmission.

Results for POH compression with image compression methods for which there are low-power codec implementations available, such as BPG (HEVC-Intra), JP2K-I and JPEG, and the simple decodable direct quantization (PCM) were analyzed to see whether they satisfy the design requirements. Flat quantization of transform coefficients in BPG and JPEG were shown to provide better results compared to the default quantization matrix. Phase unwrapping, as a pre-processing step to smooth phase-only holograms, was found to have a net negative effect on the compression performance since the increase in compressibility was lower than the increase in dynamic range of phase samples (and thus, raw data size). PCM performs the best among all codecs, attaining acceptable quality at 3 bpp, reducing the uncompressed 1.85 Gbit/s rate to ~700Mbit/s, which is still too high for reliable wireless transmission from the server to HMHD. To this effect, a new versatile PCM-POH codec with SLM-dynamic-range-aware progressive quantization and per-pixel rate-distortion control features is proposed. The new codec exploits the redundancy in the display hardware and in the content to cut the bitrate down to less than 200 Mbit/s, rendering use of current wireless communication technologies possible to transmit display-ready POH from the server to HMHD. In conclusion, this paper demonstrates that an interactive, multi-user and quality-ensured HMHD-based augmented reality system built with commercially available components is feasible using our design methodology and the proposed light-weight codec.

## Acknowledgement

This project was funded by the European Research Council (ERC) under the European Union's Seventh Framework Program (FP7/2007-2013) / ERC advanced grant agreement (340200) that ended in December 2018.

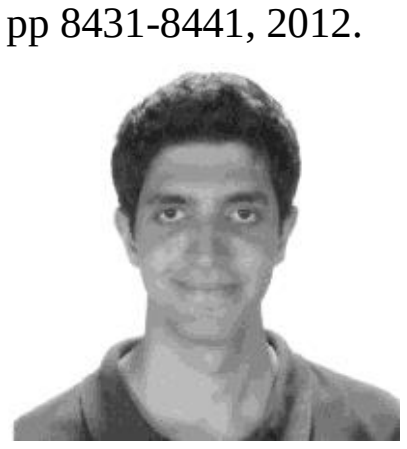

**Burak Soner** (M’19) received his BSc in Mechatronics from Sabancı University, Istanbul, Turkey, in 2014. Until 2016, he worked on power electronics at the Sabancı University Microsystems Lab as a researcher and on automotive embedded control/safety systems at AVL List GmbH as a Systems Engineer. He started his PhD in EE at Koç University (KU), Istanbul, Turkey on September 2016, and worked on real-time computation and compression for holographic displays at CY Vision, San Jose, CA and at the Optical Microsystems Laboratory in KU until 2019. He joined the Wireless Networks Laboratory in KU on January 2019 and is currently working on vehicular communications and navigation. His specific research interests are optical wireless communication and communication-based positioning/navigation systems for ground and aerial vehicles.

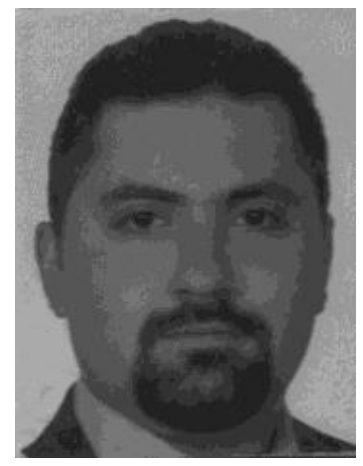

**Erdem Ulusoy** received his BS and PhD degrees both from Bilkent University, Department of Electrical and Electronics Engineering, in 2004 and 2012, respectively. During his PhD, he was a member of the EC funded 3DTV Network of Excellence project. Since 2014, he has been a member of the ERC Advanced Grant Project WEAR3D as a Research Assistant Professor in the Optical Microsystems Laboratory in Koç University, Istanbul, Turkey. His research interests include Fourier Optics, Computer Generated Holography, Light Field Synthesis with Spatial Light Modulators and Signal and Image Processing Problems in Diffraction and Holography.

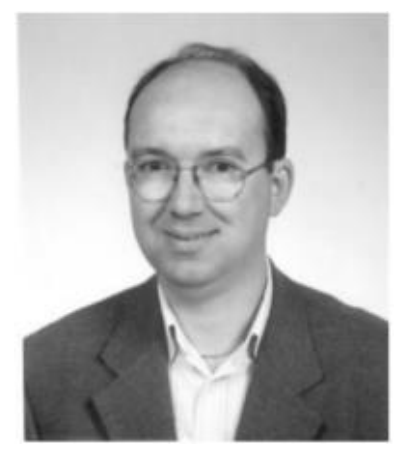

**A. Murat Tekalp** (S’80-M’84-SM’91-F’03) received Ph.D. degree in Electrical, Computer, and Systems Engineering from Rensselaer Polytechnic Institute (RPI), Troy, New York, in 1984, He has been with Eastman Kodak Company, Rochester, New York, from 1984 to 1987, and with the University of Rochester, Rochester, New York, from 1987 to 2005, where he was promoted to Distinguished University Professor. He is currently Professor at Koc University, Istanbul, Turkey. He served as Dean of Engineering between 2010-2013. His research interests are in digital image and video processing, including video compression and streaming, motion-compensated filtering, super-resolution, video segmentation and tracking, 3D video processing, and video networking.

He is a member of Academia Europaea. He chaired the IEEE SPS Technical Committee on Image and Multidimensional Signal Processing (1996-1997). He was the Editor-in-Chief of Signal Processing: Image Communication published by Elsevier (1999-2010). He served as the General Chair of IEEE Int. Conf. on Image Processing (ICIP) at Rochester, NY in 2002. He was on the editorial board of the IEEE Signal Processing Magazine (2007-2010). He served as a member of ERC Advanced Grants evaluation panel 2009-2015. He is serving in the Editorial Board of Proceedings of the IEEE since 2014. Dr. Tekalp authored the Prentice Hall book Digital Video Processing (1995), a completely rewritten second edition of which is published in 2015.

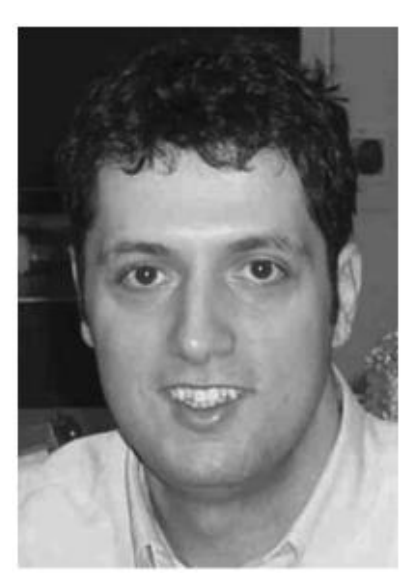

**Hakan Urey** (M’92 – SM’09) received the B.S. degree in electrical engineering from Middle East Technical University, Ankara, Turkey, in 1992, and the M.S. and Ph.D. degrees in electrical engineering from the Georgia Institute of Technology, Atlanta, GA, USA, in 1996 and 1997, respectively.

After completing the Ph.D. degree, he joined Microvision, Inc., Seattle, WA, USA, as a Research Engineer, and he played a key role in the development of scanning display technologies. He was the Principal System Engineer when he left Microvision, Inc. in 2001 to join the College of Engineering, Koc¸ University, Istanbul, Turkey, where he established the Optical Microsystems Laboratory. He is currently a Professor of electrical engineering at Koc¸ University. He has published more than 70 journal and more than 150 conference papers, six edited books, and four book chapters, and he has about 60 issued and pending patents, which have been licensed to industry for commercialization and resulted in 4 spin-off companies. His research interests include micro-electromechanical systems, micro-optics, micro-opto-electromechanical systems design, and laser-based 2-D/3-D display and imaging systems.

Dr. Urey is a Member of the SPIE, OSA, and the IEEE Photonics Society, and the Vice-President of the Turkey Chapter of the IEEE Photonics Society. He received the Werner von Siemens Faculty Excellence Award from Koc¸ University in 2006, the Distinguished Young Scientist Award from the Turkish Academy of Sciences in 2007, the Encouragement Award from The Scientific and Technological Research Council of Turkey in 2009, the Outstanding Faculty of the Year Award from Koc¸ University in 2013, and the European Research Council Advanced Investigator Grant in 2013.